\documentclass[twocolumn]{aastex61}
\received{July 1, 2016}
\revised{September 27, 2016}
\accepted{\today}
\submitjournal{ApJ}

\shorttitle{SDSS\,J141324.27+530527.0: A New ``Changing-Look'' Quasar}
\shortauthors{Wang et al.}

\begin{document}

\title{Identification of SDSS\,J141324.27+530527.0 as A New ``Changing-Look'' Quasar with a ``Turn-on'' Transition}

\correspondingauthor{J. Wang}
\email{wj@bao.ac.cn}
\correspondingauthor{D. W. Xu}
\email{dwxu@bao.ac.cn}

\author{J. Wang}
\affil{Key Laboratory of Space Astronomy and Technology, National Astronomical Observatories, Chinese Academy of Sciences, Beijing
100012, China}
\affil{School of Astronomy and Space Science, University of Chinese Academy of Sciences, Beijing, China}

\author{D. W. Xu}
\affil{Key Laboratory of Space Astronomy and Technology, National Astronomical Observatories, Chinese Academy of Sciences, Beijing
100012, China}
\affil{School of Astronomy and Space Science, University of Chinese Academy of Sciences, Beijing, China}

\author{J. Y. Wei}
\affiliation{Key Laboratory of Space Astronomy and Technology, National Astronomical Observatories, Chinese Academy of Sciences, Beijing
100012, China}
\affiliation{School of Astronomy and Space Science, University of Chinese Academy of Sciences, Beijing, China}



\begin{abstract}
We here report an identification of SDSS\,J141324+530527.0 (SBS\,1411+533) at $z=0.456344$ as a new ``changing-look'' quasar with 
a ``turn-on'' spectral type transition from Type-1.9/2 to Type-1 within a rest frame time scale of 1-10 yr
by a comparison of our new spectroscopic observation and the Sloan Digital Sky Survey (SDSS) archive data base. 
The SDSS DR7 spectrum taken in 2003 is dominated by a starlight emission from host galaxies redward 
of the Balmer limit, and has non-detectable broad H$\beta$ line.  
The new spectrum taken by us on June 1st, 2017 and SDSS DR14 spectrum taken on May 29, 2017 indicate that the object is of a 
typical quasar spectrum with a blue continuum and strong Balmer broad emission lines. In addition, an intermediate spectral type can be 
identified in the SDSS DR13 spectrum taken in 2015. The invariability of the line wing of \ion{Mg}{2}$\lambda2800$ emission and time scale argument
(The invariability of [\ion{O}{3}]$\lambda$5007 line blue asymmetry)
suggests that a variation of obscuration (an accelerating outflow) is not a favorable scenario. The time scale argument allows us to believe the type transition
is possibly caused by either a viscous radial inflow or a disk instability around a $\sim5-9\times10^{7}M_\odot$ black hole.   
\end{abstract}




\section{Introduction} \label{sec:intro}

Based on their observational properties, the zoo of active galactic nuclei (AGNs) are 
classified into various types. These types can be traditionally understood by 
the widely accepted unified model due to either the orientation effect of the 
dust torus (see Antonucci 1993 for a review), or the viewing angle of the jet with respect to the line-of-sight of
an observer (Urry \& Padovani 1995). The so-called Type-1 AGNs whose spectra show both 
broad (FWHM$>1000\ \mathrm{km\ s^{-1}}$) and 
narrow (FWHM$\sim10^2 \mathrm{km\ s^{-1}}$) Balmer emission lines are the objects associated with 
an almost face-on dust torus, while Type-2 AGNs with only narrow Balmer emission lines are 
the objects with an almost edge-on torus.    
In addition to Type-1 and 2 AGNs, there are intermediate AGN types, i.e., Type-1.5, 1.8 and 1.9
AGNs. In Type-1.5 AGNs, the broad H$\beta$ emission line is strong and there is an evident reflection in 
the H$\beta$ line profile. The objects with weak and absent broad H$\beta$ line are classified as 
Type-1.8 and 1.9 AGNs, respectively (e.g., Osterbrock \& Ferland 1996). In the context of 
the unified model, the intermediate type AGNs are explained by either partial obscuration (e.g., Stern \& Laor 2012)
or light scattering of the emission from the central engine (e.g., Antonucci \& Miller 1985). 
Besides the unified model, some previous studies argued that the Type-1 and 2 AGNs are caused by an evolution 
of central supermassive blackholes (SMBHs, e.g., Pentson \& Perez 1984; Wang \& Zhang 2007; Elitzur et al. 2014). 

With repeat spectroscopic observations, a few of AGNs are found to change their spectral types in a time scale of 
order of years, which is the so-called ``Changing-look'' AGNs (CL AGNs). Both ``turn-on'' and ``turn-off'' type transitions have 
been identified in previous studies. Some identified CL AGNs include: 
Mark\,1018, Mark\,590, 3C\,390.3, NGC\,2617, NGC\,4151, SDSS\,J015957.64+003310.5, SDSS\,J101152.98+544206.4 and
SDSS\,J155440.25+362952.0 (e.g., 
McElroy et al. 2016; Shappee et al. 2014; Shapovalova et al. 2010; LaMass et al. 2015; Runnoe et al. 2016; Gezari et al. 2017). 
Yang et al. (2017)
recently identified 21 new CL AGNs within a redshift range from 0.08 to 0.58 from either repeat spectroscopic observations 
of both Sloan Digital Sky Survey and Large Sky Area Multi-Object Fiber Spectroscopic Telescope (LAMOST) or 
photometric variations followed by new spectroscopic observations.
Additional two and 10 new CL quasars have been identified by Ruan et al. (2016) and Macleod et al. (2016), respectively, 
through the similar methods. 

The origin of CL AGNs is still under debate. There are several possible explanations: 1) a variation of the 
obscuration if the torus has a patchy configuration (e.g., Elitzur 2012); 2) a variation of SMBH accretion 
rate resulted from the secular evolution or instability (e.g., Elitzur et al. 2014; Gezari et al. 2017; 
Sheng et al. 2017; Yang et al. 2017);
3) an accelerating outflow launched from the 
central SMBH, which is adopted to understand the type transition occurred in NGC\,4151 (e.g., Shapovalova et al. 2010); and 4)
a tidal disruption event (TDE) caused by an accretion of the material of a star disrupted by the gravity of a central SMBH (e.g., 
Merloni et al. 2015; Blanchard et al. 2017).     

The study of type transition phenomenon in AGNs is of particular importance.
At first, in addition to as a challenge of the widely accepted AGN's
unified model, the ``turn-on'' phenomenon 
enables us to study the accretion physics around the central SMBHs. Secondly, both ``turn-off'' and ``turn-on''  phenomena provide us 
an ideal case for studying the host galaxy of a luminous AGN whose light from the host galaxy is overwhelmed by the 
luminous radiation emitted from the central accretion disk, which is necessary for studying the coevolution of SMBH and 
its host galaxy, especially at high redshift (see a recent review in Heckman \& Best 2014). Finally, the frequency of the 
``turn-off'' phenomenon has been suggested to restrict the life time of AGNs (Martini \& Schneider 2003).

In this paper, we report SDSS\,J141324.27+530527.0 (SBS\,1411+533) as a new CL quasar with a ``turn-on'' type transition in a rest frame time scale of $\sim1-10$ years. 
The paper is organized as follows. Section 2 describes our new spectroscopic observation and identification of the 
type transition phenomenon. The spectral analysis is presented in Section 3. A discussion and conclusion are given in 
the last section. A $\Lambda$CDM cosmology with
parameters $H_0=70\ \mathrm{km\ s^{-1}\ Mpc^{-1}}$, $\Omega_{\mathrm{m}}=0.3$ and $\Omega_{\Lambda}=0.7$ 
(Spergel et al. 2003) is adopted throughout the paper.  

\section{Observation and Identification} \label{sec:style}

The ``turn-on'' type transition of SDSS\,J141324.27+530527.0 (with a brightness of $r'=19.87$ at $z=0.456344$)
was serendipitously discovered when we performed a spectroscopic re-observation on 
a sample of quasars at $z\sim0.5$ with hybrid spectroscopic properties selected from 
SDSS Data Release 6 (Wang \& Wei 2009 and references therein) to study the coevolution issue of SMBHs and
their host galaxies. 
The spectra of these hybrid quasars taken by SDSS DR6/7 are dominated by a starlight emission from host galaxies redward 
of the Balmer limit and by an evident \ion{Mg}{2}$\lambda2800$ emission at the blue end.

\subsection{New P200 Spectroscopic Observation} \label{subsec:tables}

The new spectroscopic observation was carried out with Palomar observatory Hale 5m (P200) telescope on June 1st, 2017. 
Both blue and red cameras of DoubleSpec spectrograph (DBSP) were used in our observation. 
A grating with 600$\mathrm{line\ mm^{-1}}$ along with a slit of 1$\symbol{125}$ oriented
in the south–north direction is adopted for both cameras, which 
provides us a spectral resolution $\sim3\AA$ that is measured from the telluric emission lines and comparison arcs.  
This spectral resolution is comparable with the previous observations taken by SDSS.
In order to cover both \ion{Mg}{2}$\lambda2800$ emission line and 4000\AA\ break for the objects listed in our sample, 
the blazed wavelength was fixed at 5000\AA\ for the blue camera. The blazed wavelength was 
8500\AA\ for the red one to cover the H$\alpha$ emission line.   
In order to enhance the signal-to-noise ratio and to eliminate the contamination of
cosmic rays easily, the object was observed with 7 frames with an exposure time of 1200s for each frame. 
The average airmass and seeing are 1.2 and 1.0\symbol{125}, respectively, during the observation. 
The wavelength calibration was carried out
by the iron-argon comparison arc for the blue camera and by the helium–neon–argon comparison arc 
for the red one. The observed flux was calibrated by the Kitt
Peak National Observatory (KPNO) standard stars BD\,284211 and Feige\,34
(Massey et al. 1988).

\subsection{Data Reduction}

The two-dimensional spectra in both cameras were reduced by the standard
procedures through the IRAF package\footnote{IRAF is distributed by the National Optical Astronomical Observatories,
which is operated by the Association of Universities for Research in
Astronomy, Inc., under cooperative agreement with the National Science
Foundation.} including bias subtraction,
flat-field correction, and image combination along with cosmic-ray removal before the
extraction of the one-dimensional spectra. The extracted spectra in both cameras
were then calibrated in wavelength and flux by the corresponding
comparison arc and standards. The A-band telluric feature around $\lambda\lambda$7600-7630 due to 
$\mathrm{O_2}$ molecules was removed from the observed spectrum by the standard.
The Galactic extinction was corrected by the extinction magnitude of $A_\mathrm{V}=0.023$ (Schlafly \& Finkbeiner 2011) 
taken from the NASA/IAPC Extragalactic Database (NED), assuming the $R_\mathrm{V} = 3.1$ extinction law of our
Galaxy (Cardelli et al. 1989). 
The spectrum was then transformed to the rest frame, along with the correction of the relativity effect
on the flux, according to its redshift.

\subsection{Identification of  A ``Turn-on'' Type Transition}

Figure 1 shows the rest-frame P200 spectrum of the object, along with the three previous spectra taken by SDSS at different epochs. 
As an additional illustration, Figure 2 compares the four spectra for the H$\alpha$, H$\beta$ and H$\gamma$ emission 
lines, after the AGN's continuum and underlying starlight emission are removed (see Section 3.1 for the details).    
One can see from the comparison in both figures that the P200 spectrum is highly consistent with the SDSS DR14 spectrum taken on May 29, 2017.
Both spectra show that the object can be classified as a typical Type-1 AGNs (e.g., Vanden Berk et al. 2001) with dominant AGN's continuum at
the blue end and evident broad Balmer emission lines, especially the H$\beta$ and H$\gamma$ broad lines. 
In contrast to the two spectra taken in 2017, the spectrum taken by SDSS DR6/7 at May 2nd, 2003 shows not only quite weak AGN's continuum, but also 
an absence of H$\beta$ and H$\gamma$ broad emission lines, which strongly suggests that the object has an either Seyfert 2 or 1.9-like spectrum.    
In addition, compared with the 2017 and 2003 spectra, an intermediate spectral type can be identified for the  
SDSS DR13 spectrum taken on Oct 08, 2015. In the 2015 spectrum, the 4000\AA\ features are clearly diluted by the increased AGN's continuum, and
there is, if any, a considerably weak H$\beta$ broad emission line.

\begin{figure}[ht!]
\plotone{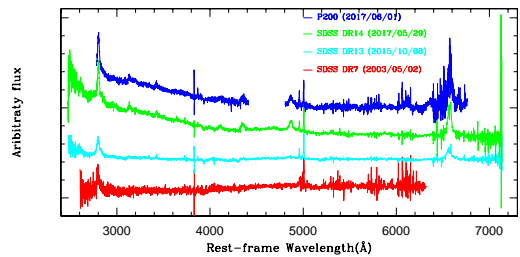}
\caption{A comparison of the new spectrum taken by P200 telescope on Jun 1st, 2017 and the spectra extracted from 
the SDSS archive data bases. All the spectra are transformed to the rest frame, and are shifted vertically by an
arbitrary amount for visibility.}
\end{figure}

\begin{figure}[ht!]
\plotone{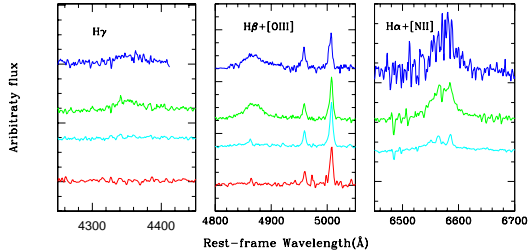}
\caption{A comparison of the Balmer emission line profiles, after the AGN's continuum and underlying host galaxy emission are 
removed (see Section 3.1 for the details). The symbol is the same as in Figure 1. Again,
all the spectra are shifted vertically by an arbitrary amount for visibility.}
\end{figure}

In summary, with the four spectra taken at different epochs, we clearly see a ``turn-on'' type transition occurring in SDSS\,J141324+530527, in which 
its spectral type changes from Type-2/1.9 to Type-1 within a  rest frame time scale of 1-10 years.

\section{Spectral Analysis}

In this section, we perform a spectral analysis on the spectra taken at the four different epochs to 
shed a light on the ``turn-on'' transition occurring in the object.

\subsection{AGN's Continuum and Stellar Features Removal}

At the beginning, the SDSS DR7 spectrum is modeled by a linear combination of a powerlaw from the central 
AGN and the emission from the underlying host galaxy (e.g., LaMassa et al. 2015; Ruan et al. 2016), because the continuum of the 
spectrum is dominated by the emission from the host galaxy, which enables us to determine the underlying spectrum of 
the host easily. The host galaxy template with an age ranging from 5Myr to 10Gyr is extracted from the single stellar population (SSP)
spectral library given by Bruzual \& Charlot (2003). We simply fix the line width of the templates to be twice of the SDSS spectral resolution,
because of the poor signal-to-noise ratio of the stellar features. We perform a $\chi^2$ minimization iteratively
over the whole spectroscopic wavelength range in the observe-frame, except for
the regions with known emission lines ((e.g., H$\beta$, H$\gamma$, H$\delta$, [\ion{O}{3}]$\lambda\lambda$4959, 5007, 
[\ion{O}{2}]$\lambda$3727, [\ion{Ne}{3}]$\lambda$3869, [\ion{Ne}{5}]$\lambda$3426, and \ion{Mg}{2}$\lambda$2800)). 
The contribution from both optical and ultraviolet \ion{Fe}{2} complex is ignored because they are too weak to be modeled.
The continuum removal for the SDSS DR7 spectrum is illustrated in the bottom panel in Figure 3. The best fit shows that the underlying host galaxy
emission is dominant by a SSP spectrum with an age roughly at 5Gyr.

With the determined underlying host galaxy spectrum, we then model the three spectra taken in 2015 and 2017 by using more
complicated models, which are presented in Figure 3.  Each of the three spectra is modeled by a linear combination of the following components:
(1) an AGN's powerlaw continuum, (2) the host galaxy spectrum determined from the SDSS DR7 spectrum,
in which the level of the host spectrum is not fixed in the modeling to account of the light loss issue due to different 
fiber (slit) widths and seeing values (e.g., Ruan et al. 2016), (3) a template of both high-order Balmer
emission lines and a Balmer continuum from the broad-line region (BLR).  and (4) an empirical template of
ultraviolet Fe II complex. The optical \ion{Fe}{2} complex is again ignored in the modeling because of its weakness in all the 
spectra. The theoretical template (Bruhweiler \& Verner 2008) of the ultraviolet \ion{Fe}{2} complex giving the best
fit to the observed I\,ZW1 spectrum is used in our continuum modeling. The line width of the template is fixed in advance 
to be that of the broad component of H$\beta$, which is determined by our line profile modeling (see below). 

The template of the Balmer continuum $f_\lambda^{\mathrm{BC}}$ is built from the emission from a partially optically 
thick cloud with an electron temperature of $T_e=1\times10^4$K by following Dietrich et al. (2002, see also in Grandi 1982 and Malkan \& Sargent 1982):
\begin{equation}
  f_\lambda^{\mathrm{BC}}=f_\lambda^{\mathrm{BE}}B_\lambda(T_e)(1-e^{-\tau})\ \lambda\leq\lambda_{\mathrm{BE}}
\end{equation}
where $f_\lambda^{\mathrm{BE}}$ is the continuum flux at the Balmer edge $\lambda_{\mathrm{BE}}=3646$\AA\ and
$B_\lambda(T)$ is the Planck function. $\tau_\lambda$ is the optical depth at wavelength $\lambda$, which is related to the one at the Balmer edge
$\tau_{\mathrm{BE}}$ as $\tau_\lambda=\tau_{\mathrm{BE}}(\lambda/\lambda_{\mathrm{BE}})^3$. A typical value of
$\tau_{\mathrm{BE}}=0.5$ is adopted in the current modeling.

The high-order Balmer lines (i.e., H7-H50) are modeled by the case B recombination model with an electron temperature
of $T_e=1.5\times10^4$K and an electron density of $n_e=10^{8-10}\ \mathrm{cm^{-3}}$ (Storey \& Hummer 1995).
The widths of these high-order Balmer lines are, again, 
determined from the line profile modeling of the H$\beta$ broad emission (see below).

\begin{figure}[ht!]
\plotone{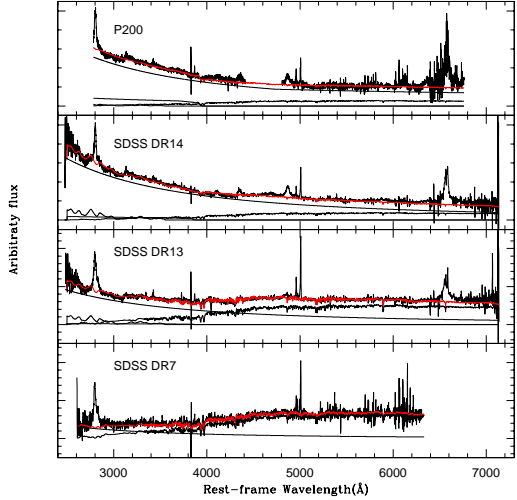}
\caption{Illustration of the modeling and removal of the continuum for P200, SDSS DR14 and SDSS DR13 spectra (from top to bottom). 
In each panel, the top heavy
curve shows the observed rest-frame spectrum overplotted by the best fitted continuum by the red curve. The underneath light curves show
the individual components used in the modeling. }
\end{figure}

\subsection{Line Profile Modeling}

After removing the underlying continuum, the emission-line profiles are modeled on each emission-line isolated spectrum
for both H$\alpha$ and H$\beta$ regions by the SPECFIT task
(Kriss 1994) in the IRAF package. The profile modeling of the H$\alpha$ region is abandoned for the P200 spectrum because of 
its poor S/N ratio at the red end. 
The flux of [\ion{O}{3}]$\lambda5007$ emission line
of the SDSS DR7 spectrum is obtained from a direct integration.
In the profile modeling, each [\ion{O}{3}]$\lambda$5007 line profile is modeled by a sum of two Gaussian functions.  
In addition to the narrow component with a width of several hundreds of $\mathrm{km\ s^{-1}}$, a broad and blue shifted component is
usually required to reproduce the observed [\ion{O}{3}] line profile in AGNs 
(e.g., Boroson 2005; Harrison et al. 2014; Zhang et al. 2013; Woo et al. 2017; Wang et al. 2011, 2017
and references therein).  
The line flux ratios of the [\ion{O}{3}]$\lambda\lambda$4959, 5007 and
[\ion{N}{2}]$\lambda\lambda$6548, 6583 doublets are fixed to their theoretical values. Two Gaussian profiles, a broad and a narrow components,
are required to adequately reproduce the H$\alpha$ line profiles in all the three spectra taken in 2015 and 2017. 
One broad component is sufficient to reproduce the observed H$\beta$ line profiles adequately in both spectra taken in 2017. 
The line modelings are presented in the left and right panels of Figure 4 for the H$\beta$ and H$\alpha$ regions,
respectively. The results of the spectral modeling are listed in Table 1.  No intrinsic extinction correction is applied 
to  all the derived line fluxes both because the traditionally used method based on the Balmer decrement of narrow emission lines
is unavailable for the current spectra and because the Balmer decrement obtained from the broad emission lines is
H$\alpha/$H$\beta=2.46\pm0.26$ that is close to the standard case B recombination (e.g., Dong et al. 2008).
All the errors reported in the table correspond to the
1$\sigma$ significance level after taking into account the proper error
propagation.  

\begin{figure}[ht!]
\plotone{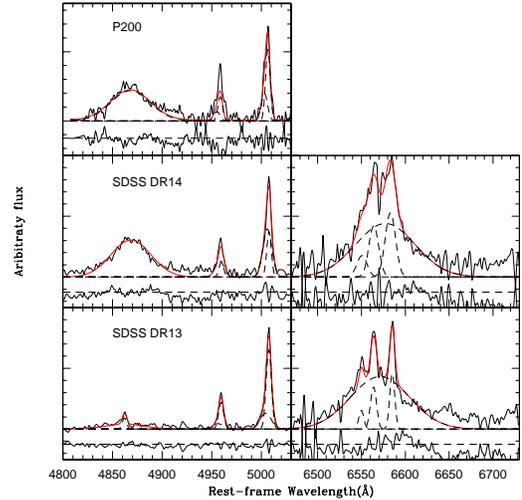}
\caption{\it Left panels: \rm Line profile modelings by a linear combination of a set of Gaussian functions for
the H$\beta$ region, in which the modeled continuum has already been removed from the original observed spectrum. The observed
and modeled line profiles are plotted by black and red solid lines, respectively. Each Gaussian function is shown by a dashed
line. The sub-panel underneath the line spectrum presents the residuals between the observed and modeled profiles. \it Right panels:
\rm The same as the middle ones but for H$\alpha$ region.}
\end{figure}

\begin{table*}[h!]
\renewcommand{\thetable}{\arabic{table}}
\centering
\caption{Spectral measurements and analysis.} \label{tab:decimal}
\begin{tabular}{ccccc}
\tablewidth{0pt}
\hline
\hline
Property & SDSS DR7 & SDSS DR13 & SDSS DR14 & P200 \\ 
\hline
epoch    & 2003/05/02 & 2015/10/08 & 2017/05/29 & 2017/06/01 \\  
\hline
AGN type & 1.9/2.0  &  1.8/1.9  &  1.0  & 1.0 \\
\hline
F([OIII]$\lambda5007$) & $0.90\pm0.07$  &  $1.22\pm0.05$ & $1.34\pm0.04$ & $0.93\pm0.20$ \\
$\mathrm{10^{-15}\ erg\ s^{-1}\ cm^{-2}}$ &  & & & \\
\hline
F(H$\beta_\mathrm{b}$) & \dotfill &  $0.39\pm0.04$ & $3.05\pm0.14$ &  $2.00\pm0.18$\\
$\mathrm{10^{-15}\ erg\ s^{-1}\ cm^{-2}}$ &  & & & \\
FWHM(H$\beta_\mathrm{b}$) &  \dotfill & $2780\pm220$ &  $2900\pm160$ &  $2900\pm60$\\
$\mathrm{km\ s^{-1}}$ &  & & & \\
\hline
F(H$\alpha_\mathrm{b}$) & \dotfill &  $3.64\pm0.19$ & $7.49\pm0.70$ & \dotfill \\
$\mathrm{10^{-15}\ erg\ s^{-1}\ cm^{-2}}$ &  & & & \\
FWHM(H$\alpha_\mathrm{b}$) &  \dotfill & $3610\pm230$ &  $3640\pm220$ & \dotfill \\
$\mathrm{km\ s^{-1}}$ &  & & & \\
\hline
$M_{\mathrm{BH}}/M_\odot$\tablenotemark{*} & \dotfill &  \dotfill & $6.8\times10^7$ & $6.1\times10^7$\\
                          & \dotfill & $7.0\times10^7$ &  $9.5\times10^7$ & \dotfill \\
\hline
$L_{\mathrm{bol}}$\tablenotemark{*} & \dotfill &  \dotfill & $1.4\times10^{45}$ & $1.2\times10^{45}$\\
$\mathrm{erg\ s^{-1}}$   & \dotfill & $5.7\times10^{44}$ &  $9.8\times10^{44}$ & \dotfill \\
\hline
$L_{\mathrm{bol}}/L_{\mathrm{Edd}}$\tablenotemark{*} & \dotfill &  \dotfill & 0.16 & 0.16\\
                                    & \dotfill & 0.06 & 0.08 & \dotfill \\
\hline
\end{tabular}
\tablenotetext{*}{For each property, the first line is based on the measurements of H$\beta$ broad emission
lines, and the second line on that of H$\alpha$. }
\end{table*}

\subsection{[\ion{O}{3}]$\lambda5007$ and \ion{Mg}{2}$\lambda2800$ Emission Line Profile}

The left panel in Figure 5 compares the [\ion{O}{3}]$\lambda5007$ emission line profiles taken at the four different epochs.  
In the comparison, the rest-frame line profile taken by P200 is convolved with a Gaussian function with a width of
$\sigma=\sqrt{\sigma_{\mathrm{SDSS}}^2-\sigma_{\mathrm{P200}}^2}/(1+z)$ to match the instrumental resolution of the P200 spectrum to 
that of the SDSS spectra, where $\sigma_{\mathrm{SDSS}}$ and $\sigma_{\mathrm{P200}}$ are the instrumental resolution at the observer frame,
respectively, and $z$ is the redshift of the object. The comparison clearly indicates that there is no detectable variation of 
the [\ion{O}{3}] line profile with a time scale of a dozen years. In fact, an extremely high consistence of the [\ion{O}{3}] line profile 
can be found for the SDSS 2015 and 2017 spectra in which the object changes its spectral type from type 1.8/1.9 to type 1.    
 
Not as the statement in Gezari et al. (2017) for a lack of strong variation of \ion{Mg}{2}$\lambda2800$ line emission for 
CL AGNs, a dramatic line profile variation can be identified for the \ion{Mg}{2} emission line 
in the right panel of Figure 5. When the spectral type changes from 1.8/1.9 to 1.0, the \ion{Mg}{2} line core emission increases significantly, although 
its high velocity wings  are still invariable.

\begin{figure}[ht!]
\plotone{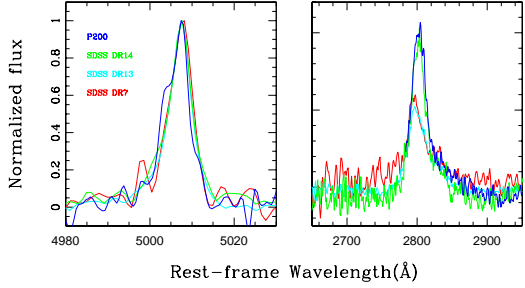}
\caption{\it Left panel: \rm A comparison of the normalized [\ion{O}{3}]$\lambda5007$ emission line profile. The spectral resolution of the P200 spectrum is 
scaled to that of SDSS spectra by a convolution with a Gaussian function. See Section 3.2 for the details. The symbols are the same as in Figure 1.
\it Right panel: \rm the same as the left one but for \ion{Mg}{2}$\lambda2800$ emission line.}
\end{figure}

\subsection{Analysis}

With the line profile modelings, we start here to estimate 
both SMBH viral mass ($M_{\mathrm{BH}}$) and Eddington ratio $L/L_{\mathrm{Edd}}$ (where 
$L_{\mathrm{Edd}} = 1.26\times10^{38}M_{\mathrm{BH}}/M_\odot\ \mathrm{erg\ s^{-1}}$ is the Eddington
luminosity) that are critical parameters
describing AGN phenomena (e.g., Shen \& Ho 2014 and references therein), basing upon the several calibrated relationships.
These calibrations enable us to estimate $M_{\mathrm{BH}}$ and  $L/L_{\mathrm{Edd}}$ from different
broad emission lines in a single epoch spectroscopy (e.g., Wu et al. 2004), thanks to the great progress 
made in the reverberation mapping technique (e.g., Kaspi et al. 2000, 2005; Peterson \&
Bentz 2006; Marziani \& Sulentic 2012; Peterson 2014; Wang et al. 2014; Du et al. 2014, 2015).

For H$\beta$ broad emission line, we use the calibration in Vestergaard \& Peterson (2006) 
\begin{equation}
  M_{\mathrm{BH}}=10^{6.67}\bigg(\frac{L_{\mathrm{H\beta}}}{10^{42}\ \mathrm{erg\ s^{-1}}}\bigg)^{0.63}\bigg(\frac{\mathrm{FWHM(H\beta)}}{1000\ \mathrm{km\ ^{-1}}}
\bigg)^2M_\odot
\end{equation}
to obtain an estimation of $M_{\mathrm{BH}}$. The bolometric luminosity $L_{\mathrm{bol}}$ is then estimate from the usually used
bolometric correction $L_{\mathrm{bol}}=9\lambda L_\lambda(5100\AA)$ (Kaspi et al. 2000), where 
$L_{\lambda}(5100\AA)$ is the AGN's specific continuum luminosity at 5100\AA\ that can be inferred from H$\beta$ broad line luminosity through the calibration given in Greene \& Ho
(2005)
\begin{equation}
 \lambda L_{\lambda}(5100\AA)=7.31\times10^{43}\bigg(\frac{L_{\mathrm{H\beta}}}{10^{42}\ \mathrm{erg\ s^{-1}}}\bigg)^{0.883}\ \mathrm{erg\ s^{-1}}
\end{equation}

In the case of H$\alpha$ broad emission line,  $M_{\mathrm{BH}}$ is estimated from the calibration provided in Greene \& Ho (2007)
\begin{equation}
   M_{\mathrm{BH}}=3.0\times10^6\bigg(\frac{L_{\mathrm{H\alpha}}}{10^{42}\ \mathrm{erg\ s^{-1}}}\bigg)^{0.45}
\bigg(\frac{\mathrm{FWHM(H\alpha)}}{1000\ \mathrm{km\ s^{-1}}}\bigg)^{2.06} M_\odot
\end{equation}
and the luminosity at 5100\AA\ from the $L_\lambda(5100)-L_{\mathrm{H\alpha}}$ relationship in Greene \& Ho (2005)
\begin{equation}
   \lambda L_{\lambda}(5100\AA)=2.4\times10^{43}\bigg(\frac{L_{\mathrm{H\alpha}}}{10^{42}\ \mathrm{erg\ s^{-1}}}\bigg)^{0.86}\ \mathrm{erg\ s^{-1}}
\end{equation}

The estimated $M_{\mathrm{BH}}$, bolometric luminosity $L_{\mathrm{bol}}$ and $L/L_{\mathrm{Edd}}$ are tabulated in Table 1. 
In the estimation, the used H$\alpha$ and H$\beta$ line fluxes are calibrated by a constant 
total flux of [\ion{O}{3}]$\lambda5007$ of the SDSS DR14 spectrum, because of the invariability of the [\ion{O}{3}] line profile 
shown in Section 3.2. The resulted $L_{\mathrm{bol}}$ ($>10^{44}\ \mathrm{erg\ s^{-1}}$) enables us to classify SDSS\,J141324+530527 as a 
quasar with a $M_{\mathrm{BH}}$ of $5-9\times10^7M_\odot$ at a moderate $L/L_{\mathrm{Edd}}\sim0.1$.


\section{Conclusion and Discussion} \label{sec:pubcharge}

By comparing the new optical spectrum taken by P200 telescope on June 1st, 2017 and three previous spectra taken by SDSS, 
we report SDSS\,J141324+530527 is a new CL quasar ($z=0.456344$ and $M_{\mathrm{BH}}\sim5-9\times10^7M_\odot$) with 
a ``turn-on'' type transition from Type-2/1.9 to Type-1 within a rest frame time scale of 1-10 years. 

We argue that the variation of obscuration is a disfavored explanation for the type transition observed in 
the object based on the time scale argument. A crossing time can be estimated from the Equation 4 in LaMassa et al. (2015) 
for the obscuration material orbiting outside the BLR on a circular, Keplerian orbit as
\begin{equation}
  t_{\mathrm{cross}}=0.07\bigg(\frac{r_{\mathrm{orb}}}{1\mathrm{ld}}\bigg)^{3/2}\sin^{-1}\bigg(\frac{r_\mathrm{src}}{r_\mathrm{orb}}\bigg)
\bigg(\frac{M_{\mathrm{BH}}}{10^8M_\odot}\bigg)^{-1/2}
\end{equation}
where $r_{\mathrm{orb}}$ and $r_{\mathrm{src}}$ are the orbital radius of the obscuration material and 
the true size of the BLR, respectively. The characteristic radius of BLR, $R_{\mathrm{BLR}}$, is estimated to be $\sim47$ days from the H$\beta$ line 
luminosity through a combination of Equation (2) and the radius-luminosity relationship 
$\log(R_{\mathrm{BLR}}/\mathrm{1ld})=1.559+0.549\log(\lambda L_\lambda(5100\AA)/10^{44}\ \mathrm{erg\ s^{-1}})$ 
given in Bentz et al. (2013), which yields a $t_{\mathrm{cross}}>42$yr when $M_{\mathrm{BH}}=7\times10^7M_\odot$ and $r_{\mathrm{orb}}=R_{\mathrm{BLR}}$ are
adopted. This crossing time scale is obviously larger than the type transition time observed in the object.   
The disapproval of the obscuration scenario is further supported by the observed line profile variation of the \ion{Mg}{2}$\lambda2800$
emission line profile. For BLR gas in a circular, Keplerian orbit around the central SMBH, the line wings are produced by the emission from 
the high velocity gas at inner part of BLR, and the line core the low velocity gas at outer part. This configuration suggests that 
the strength of line wings are more sensitive to the obscuration than the line core, which is, however, in contradiction with the observed line profile 
variation.

The scenario of an accelerating outflow is also seemed to be disfavored because of the invariability of the blue asymmetry of 
the [\ion{O}{3}]$\lambda5007$ line profile, which is usually used as an assessment of the strength of outflow in AGNs (e.g.,
Wang et al. 2017 and references therein).

Another considered mechanism of AGN's spectral type transition is the TDE in which a star close enough to the SMBH is disrupted 
and about half of the material of the star is accreted by the SMBH to rapidly increase accretion rate (e.g., Rees 1988).  
However, both 2017 spectroscopic observations taken at the ``turn-on'' phase suggest that the spectra of the object more resemble to a typical AGN's spectrum rather 
than a TDE spectrum typically with strong helium emission (e.g., Gezari et al. 2012; Arcavi et al. 2014).
Follow-up observations, especially the photometry observations, are useful for determine the truth of a TDE in the object. 
The brightness of the TDE typically follows a $t^{-5/3}$ power-law decay with a time scale of
$\Delta t=0.35(M_{\mathrm{BH}}/10^7M_\odot)^{1/2}(M_\star/M_\odot)^{-1}(R_\star/R_\odot)^{3/2}\sim 1$yr (e.g., Rees 1998; Lodato \& Rossi 2011 and references therein), 
where $M_\star$ and $R_\star$ are the mass and radius of the disrupted star.    

The time scale argument allows us to argue that the viscous radial inflow is a plausible scenario for the observed type transition. 
The accretion rate of a central SMBH can change its value due to an inflow of gas in the inner parts of the accretion disk with a size of $r$ at a 
time scale of (e.g, Gezari et al. 2017)
\begin{equation}
  \tiny
  t_{\mathrm{infl}}=1300\bigg(\frac{\alpha}{0.1}\bigg)^{-1}\bigg(\frac{L/L_{\mathrm{Edd}}}{0.005}\bigg)^{-2}\bigg(\frac{\eta}{0.1}\bigg)^{2}
\bigg(\frac{r}{10r_g}\bigg)^{7/2}\bigg(\frac{M_{\mathrm{BH}}}{2\times10^8M_\odot}\bigg)\mathrm{yr}
\end{equation}
where $\alpha$ is the ``viscosity'' parameter, $\eta$ the
efficiency of converting potential energy to radiation, and $r_g$ the gravitational radius.
Adopting the typical values of $\alpha=\eta=0.1$ and $r=10r_g$ (the  typical ultraviolet emitting region of the disk)  and 
the estimations of both $M_{\mathrm{BH}}$ and $L/L_{\mathrm{Edd}}$ returns a 
$t_{\mathrm{infl}}\sim 1-5$yr, which is quite comparable to the observed type transition time scale in the object.  

The variability due to disk local thermal instability is an additional plausible scenario. According to the evolutionary $\alpha-$disk model developed
in Siemiginowska et al. (1996), the thermal time scale is 
\begin{equation}
  t_{\mathrm{th}}\sim\frac{1}{\alpha\Omega_{\mathrm{K}}}=2.7\bigg(\frac{\alpha}{0.1}\bigg)^{-1}\bigg(\frac{r}{\mathrm{10^{16}\mathrm{cm}}}\bigg)^{3/2}
\bigg(\frac{M_{\mathrm{BH}}}{10^8 M_\odot}\bigg)^{-1/2}\mathrm{yr}
\end{equation}
Taking  $M_{\mathrm{BH}}=7\times10^7M_\odot$ yields a $t_{\mathrm{th}}\sim3.2$yr, which is on agreement with the observed type transition 
time scale in the case of SDSS\,J141324+530527, although the reality of the inhomogeneous disk model is either argued against or supported by some previous 
studies (e.g., Hung et al. 2016; Ruan et al. 2014). 

In fact, the mechanism of variation of the accretion rate is found to be a favorite scenario in some CL AGNs due to the detection of 
large variation of brightness in mid-infrared (Sheng et al. 2017), because the infrared emission is
much less sensitive to dust extinction than the radiation in optical band. A similar conclusion is arrived by Yang et al. (2017) who reported a 
brightening in infrared for some CL AGNs when the AGNs turned on. In Mark\,1018, the scenario of accretion rate variation, rather than obscuration, 
is favored both because of the followed blackbody $L\sim T^4$ relation and because of the lack of hydrogen absorption in its X-ray spectrum (Husemann et al. 2016).

\acknowledgments

JW \& DWX are supported by the National Natural Science Foundation of China under grants
11473036 and 11773036. The study is supported by the National Basic Research Program of China (grant
2014CB845800) and by the Strategic Pionner Program on Space Science, Chinese Academy of Sciences (Grant
No.XDA15052600). 
This study uses the SDSS archive data that was created and distributed by the Alfred P.
Sloan Foundation, the Participating Institutions, the National Science
Foundation, and the U.S. Department of Energy Office of Science.
This research uses data obtained through the Telescope Access Program (TAP), 
which has been funded by the National Astronomical Observatories of China, the Chinese Academy of Sciences, 
and the Special Fund for Astronomy from the Ministry of Finance. Special thanks go to the staff at Palomar Observatory
for their instrumental and observational helps.
\vspace{5mm}
\facilities{Palomar Hale 5m Telescope}
\software{IRAF (Tody 1986, Tody 1993)}

\end{document}